# Design and theory of switchable linear magnetoelectricity by ferroelectricity in Type-I multiferroics

Hui-Min Zhang[1], Cheng-Ao Ji[1], Tong Zhu[2], Hongjun Xiang[3], Hiroshi Kageyama[2], Shuai Dong[1], James M. Rondinelli[4],* and Xue-Zeng Lu[1],*

[1]Key Laboratory of Quantum Materials and Devices of Ministry of Education, School of Physics, Southeast University, Nanjing 211189, China
[2]Graduate School of Engineering, Kyoto University, Nishikyo-ku, Kyoto 615-8510, Japan
[3]Key Laboratory of Computational Physical Sciences (Ministry of Education), Institute of Computational Physical Sciences, and Department of Physics, Fudan University, Shanghai 200433, China
[4]Department of Materials Science and Engineering, Northwestern University, Evanston, Illinois 60208, USA
*jrondinelli@northwestern.edu
*xuezenglu@seu.edu.cn

**Abstract**

We present a comprehensive theoretical investigation of magnetoelectric (ME) coupling mechanisms in 19 altermagnetic and 4 ferrimagnetic Type-I multiferroics using electronic band structure calculations with spin-orbit coupling, a first-principles ME response framework, and spin-space-group theory analysis. We formulate a universal scheme for realizing nonvolatile ME coupling in Type-I multiferroics, where two distinct pathways emerge, each dictated by spin-space symmetry. The first pathway is associated with switching of the spin splitting or the now familiar spin-momentum locking in reciprocal space, characteristic of some altermagnetic multiferroics that exhibit coexisting antiferromagnetism and ferroelectricity. The second pathway involves real-space magnetization switching via electric polarization reversal, characterized by switchable components of the linear ME tensor, despite the traditionally weak coupling in Type-I systems due to the independent origins of magnetism and ferroelectricity. We demonstrate that these two intrinsic ME coupling mechanisms are mutually exclusive and propose thermodynamically stable compounds for experimentation. Our findings establish general design principles for controlling robust nonvolatile ME effects in multiferroic materials.

*Introduction*—Ferroelectric perovskite oxides have been studied since the discovery of ferroelectricity in BaTiO$_3$. Many of these, such as BaTiO$_3$, PbTiO$_3$, and PMN-PT (lead magnesium niobate-lead titanate), are employed in sensors, actuators, nonvolatile memory devices, and medical imaging [1–4]. By further combining ferroelectricity with magnetism, functional multiferroic materials have created a new platform for realizing high-density and nonvolatile memory devices that operate at lower power [5–8]. Designing new high-temperature single-phase multiferroic materials with strong magnetoelectricity, however, is still challenging and ultimately necessary for realizing electric-field control of magnetism [6].

Altermagnetism, which was recently proposed for centrosymmetric materials with antiferromagnetically collinear magnetic structures, has attracted renewed interest in antiferromagnetic oxides [9–12]. Symmetry analysis shows that altermagnetism arises when any magnetic ions with collinear and antiparallel spins are not connected by inversion and nonsymmorphic translational operations. The absence of these real-space symmetries is reflected in band structure calculations, where spin-up and spin-down bands exhibit splitting along symmetry-breaking momentum paths—even in the absence of spin-orbit coupling (SOC). So far, several theoretical and computational studies have screened altermagnets in two-dimensional and three-dimensional material databases, and some non-centrosymmetric materials, such as Ca$_3$Mn$_2$O$_7$, have been identified as altermagnets [13]. Although many computational studies have focused on the symmetry requirements and magnitudes of the spin degeneracy lifting, enabling the ferroelectric control of altermagnetism in momentum space [14–16], which are sought after to access the Hall effect [e.g., the crystal Hall effect (CHE)] [17], very few have examined how nonvolatile magnetoelectricity can be realized from altermagnetism in real space to realize the electric field control of magnetism in multiferroics. We hypothesize that magnetoelectric (ME) coupling that originates in spatial and time (spin) reversal symmetries foundational to altermagnetism may help overcome the difficulties in realizing electric-field control of magnetism in single-phase type-I multiferroic materials.

Here, we performed first-principles calculations and comprehensively searched for possible ground state structures in a series of mixed-anion $A_2BX_3X'$ ($X$=O, S and $X'$=N, Cl, Br) compounds. Within this family, we formulated a design principle for finding ferroelectric materials and room-temperature multiferroism. These compounds further exhibit geometric improper ferroelectricity and altermagnetism. Our electronic band structure calculations with spin-orbit coupling (SOC), combined with a first-principles framework for computing the ME response and spin space group theory analysis, reveal universal electric polarization switching pathways with well-defined spin-group operations under SOC. One path, grounded in time-reversal symmetry, focuses on magnetic-field coupled properties such as the linear ME response and piezomagnetic effect. These phenomena are characterized by a spin group symmetry operator $[E||\{I|t\}]$, where $E$ is the identity operator in the spin-only space and $I$ and $t$ are inversion symmetry and a nonsymmorphic translational symmetry in real space, respectively. Upon polarization switching through this path, we compute a change in the ME coefficients in $A_2BX_3X'$ compounds and 14 additional altermagnetic multiferroics. This nonvolatile ME mechanism is general and extends beyond altermagnetic multiferroics. We find it explains the intrinsic switchable ME coefficients hidden in many previously studied multiferroics such as altermagnetic $BiFeO_3$ and ferrimagnetic $Zn_2FeOsO_6$.

*Structural Design and Ferroelectricity*—We focus on the Ruddlesden–Popper (RP) family of materials [18], which consists of $n$ perovskite $(ABO_3)_n$ layers alternating with rock-salt $AO$ layers stacked along the out-of-plane [001] direction with the chemical formula $A_{n+1}B_nO_{3n+1}$ [i.e., $(ABO_3)_n/AO$] (**Fig. 1**). For $n=1$, the connectivity between the perovskite layers is very weak and multi-mode coupling is often required to lift inversion [19], which makes it more difficult to find a ferroelectric phase. Nonetheless, there are predicted $n=1$ RP compounds [20–22], but many are identified only on a case-by-case basis and have yet to be experimentally synthesized. A general strategy to predict ferroelectricity and switchable electric and magnetic polarizations in $n=1$ RP compounds, based solely on structural factors like the tolerance factor found in $n=2$ materials [23], remains elusive.

Here, we computationally investigate heteroanionic $A_2BX_3X'$ compounds as shown in **Figs. 1a and 1b**. Compounds favorable for hosting multiple and ordered anions can be screened by considering the electronegativity of the ions and applying Pauling's second crystal rule (PSCR) [24–26]. Generally, out-of-plane trans $X$-$B$-$X'$ anion ordering occurs for large PSCR sums and has been observed experimentally in oxynitride $Nd_2AlO_3N$ and oxychloride $Sr_2CoO_3Cl$ [24,27]. The N (Cl) anion occupies the apical site of the $AlO_5N$ ($CoO_5Cl$) octahedron, reducing the ideal local $O_h$ symmetry to $C_{4v}$. **Figs. 1a and 1b** show an "antiparallel" long-range order of the heteroleptic octahedra in $A_2BX_3X'$ compounds, where adjacent layers exhibit antialigned trans $X$-$B$-$X'$ bonds. This tiling of the heteroleptic units cannot break inversion and produces a $P4/nmm$ phase (sans any other distortions). The $P4/nmm$ phase has been reported in both oxynitride and oxyhalide compounds at ambient and low pressures: Nd$_2$AlO$_3$N [24], Sr$_2$CoO$_3$Cl [24] Sr$_2$MnO$_3$Cl [28], Ca$_2$FeO$_3$Cl [29] and Sr$_2$BO$_3$F (B=Fe,Co) [30]. By lowering the tolerance factors, we searched for ground state phases in Ca$_2$FeX$_3$X' (X=O, S and X'=Cl, Br) and Ln$_2$FeO$_3$N (Ln=La, Tb, Lu) with tolerance factors ranging from 0.807-0.864 [27]. Our results showed that compounds with large tolerance factors (e.g., 0.864) adopted chiral nonpolar P2$_1$2$_1$2 symmetry, whereas those with small tolerance factors (e.g., 0.844) adopted achiral polar ground Pca2$_1$ states. The chalcohalides Ca$_2$FeS$_3$X' (X'=Cl, Br) and oxynitride Lu$_2$FeO$_3$N can have a polar Pca2$_1$ ground-state phase with the hybrid improper ferroelectricity (**Fig. 1c and Supplemental Fig. 3**) [27], and the polarization is along the in-plane direction perpendicular to the multi-anion ordering direction (**Fig. 1b**). Thus, the electric polarization may be reversed without the limitations imposed by the asymmetric anion arrangement.

*Magnetic Properties, SOC Effects, and Polarization Switching*—Next, we discuss the magnetic properties mainly for Ca$_2$FeS$_3$Br and establish that Ca$_2$FeS$_3$Br with Néel magnetic ordering along $S_x$ (**Fig. 2a**) is altermagnet by calculating the band structures without SOC. As shown in **Fig. 2b**, band splitting between the spin-up and spin-down manifolds is clearly seen in momentum space trajectories that do not contain symmetries connecting the two antiparallel magnetic sublattices, i.e., $\mathcal{M}_b$ and $\mathcal{M}_c$ in the spin group symmetries $[C_2||\{\mathcal{M}_b|t_1\}]$ and $[C_2||\{\mathcal{M}_c|t_2\}]$. Our DFT+U+SOC calculations show that the magnetic anisotropy (MA) is along the in-plane polarization direction, indicating that Ca$_2$FeS$_3$Br should exhibit the Pca2$_1$ magnetic space group (**Fig. 2a**).

Upon including SOC effects, we observe additional structure for band dispersions between bipolar states by simultaneously switching the electric polarization and *out-of-plane rotation* [i.e., (-P, -OOR, OOT), which we defined as the -P2 state] in Ca$_2$FeS$_3$Br (**Figs. 2c**). The band degeneracies exhibit large splitting and become symmetric from -$k$ to $k$ having P1 symmetry between the two states; for example, along the P'-Γ-P path without any magnetic group symmetries of Pca2$_1$ where P'-Γ (-$k_x$, -$k_y$, -$k_z$) path has opposite $k$ values to Γ-P ($k_x$, $k_y$, $k_z$) (**Figs. 2c**). In contrast, changing both the polarization and *in-plane tilt* [i.e., (-P, OOR, -OOT) or -P1 state] gives a band structure identical to that of original polar state [i.e., (P, OOR, OOT) or P state]; hence P=-P1≠-P2. When switching off the SOC effects, the band structure dependencies disappeared.

*Magnetoelectric Coupling Mechanisms*—The above phenomena can be explained by adopting spin space symmetries $[E||\{I|t\}]$ and $[\bar{E}||\mathcal{T}\{I|t\}]$ leading to the polarization switching from P to -P2, and P to -P1, respectively (**Fig. 3a**) Recalling an important spin-only group symmetry $[\bar{C}2||\mathcal{T}]$ for nonrelativistic bands of collinear magnets [9], it leads to $[\bar{C}2||\mathcal{T}]\mathcal{F}(s, k) = \mathcal{F}(s, -k) = \mathcal{F}(s, k)$, because $[\bar{C}2||\mathcal{T}]$ is a symmetry of the nonrelativistic collinear orderings. $\mathcal{F}$ represents the band energy as a function of spin $s$ and momentum $k$. $[\bar{C}2||\mathcal{T}]$ consists of $C2$ and $[\bar{E}||\mathcal{T}]$, which are a 180° rotation of a spin about its perpendicular direction and a spin-space inversion with a time reversal symmetry in real space, respectively. Then, without SOC, $\mathcal{F}_{-P2}(s, k) = [E||\{I|t\}]\mathcal{F}_P(s, k) = \mathcal{F}_P(s, -k)$ and $\mathcal{F}_{-P1}(s, k) = [\bar{E}||\mathcal{T}\{I|t\}]\mathcal{F}_P(s, k) = \mathcal{F}_P(-s, k)$, leading to $\mathcal{F}_{-P2}(s, k) = \mathcal{F}_P(s, k)$ and $\mathcal{F}_{-P1}(s, k) = \mathcal{F}_P(-s, k)$, respectively. Thus, the band dispersion of P, -P1 and -P2 states coincides with each other without SOC. With SOC ($S·L$) and due to $[\bar{C}2||\mathcal{T}](S·L) = -(S·L)$, $[\bar{C}2||\mathcal{T}]\mathcal{F}(s, k) = \mathcal{F}(s, -k) \neq \mathcal{F}(s, k)$. Then, $\mathcal{F}_{-P2}(s, k) = [E||\{I|t\}]\mathcal{F}_P(s, k) = \mathcal{F}_P(s, -k)$ will lead to symmetric bands from -$k$ to $k$ between P and -P2. This is further demonstrated in **Fig. 2d** upon including the spin information. For -P1, $\mathcal{F}_{-P1}(s, k) = \mathcal{F}_P(-s, k)$ still occurs (**Figs. S9f and S9g**), so the band dispersions of the P and -P1 states still coincide with each other with SOC. Although spin group symmetry was proposed to study the band structure without SOC, the validation of the well-defined spin group operations $[E||\{I|t\}]$ and $[\bar{E}||\mathcal{T}\{I|t\}]$ with SOC is important as verified by our DFT+SOC simulations. Furthermore, as we shown upon considering the Dzyaloshinskii-Moriya interaction (DMI), SOC is critical for determining the magnetic ordering relationship of the bipolar states, where as $[C2||\{I|t\}]$ symmetry fails (see **Fig. 3a**) [27].

From the above analysis, it can be seen that the electric polarization switching through $[E||\{I|t\}]$ cannot lead to spin-momentum locking on the bands, but there are important effects of $\mathcal{T}$ in real space—there is associated symmetry operation $[\bar{C}2||\mathcal{T}]$ that can also lead to $\mathcal{F}_{-P2}(s, k) = \mathcal{F}_P(s, -k)$. Previous studies consistently find switchable altermagnetism (spin-momentum locking in the Brillouin zone) by switching the ferroelectricity through the spin space symmetry $[\bar{E}||\mathcal{T}\{I|t\}]$ (**Fig. 3a**), or equivalently $[C2||\{I|t\}]$ without SOC for the recovery of $[\bar{C}2||\mathcal{T}]$ as a symmetry of the nonrelativistic collinear orderings [14–16,31]. Thus, any induced switchable Hall conductivity and related phenomena (e.g., magneto-optic Kerr effects) attributed to the switchable spins in momentum space that can be measured by angle-resolved photoemission spectroscopy (ARPES), essentially stem from $[\bar{E}||\mathcal{T}\{I|t\}]$ symmetry.

Here, we will elucidate the importance of the spin space symmetry $[E\|\{I|t\}]$, which exclusively gives rise to nonvolatile real-space ME coupling as described by the linear ME tensor, i.e., an essential ingredient for achieving electric-field ($\mathcal{E}$) control of magnetism ($\mathcal{M}$) in altermagnetic multiferroics due to the fact that both the linear ME response and altermagnetic ordering require $t\mathcal{T}$-symmetry breaking. According to an electric-magnetic enthalpy of an equilibrium state subjected to both electric and magnetic fields [32,33], there will be energy terms ($F$) written as $F = \alpha_{m\mu}u_m\mathcal{H}_\mu + \alpha_{\alpha\mu}\mathcal{E}_\alpha\mathcal{H}_\mu + \alpha_{m\alpha}u_m\mathcal{E}_\alpha$, where $\alpha_{\alpha\mu}$ and $\alpha_{m\alpha}$ are the frozen-ion ME response and Born-charge tensors, respectively, $u$, $\mathcal{H}$ and $\mathcal{E}$ are atomic displacements away from their equilibrium positions, magnetic and electric fields, respectively, $\alpha_{m\mu}$ is a relaxed-ion tensor that accounts for the atomic displacements $u_m$ induced by the magnetic field. By applying $[\bar{C}2\|\mathcal{T}]$ to $F_P$ in the P state (noting that only $\mathcal{T}$ works on the magnetic field), $[\bar{C}_2\|\mathcal{T}] F_P = \alpha_{m\mu}u_m(-\mathcal{H}_\mu) + \alpha_{\alpha\mu}\mathcal{E}_\alpha(-\mathcal{H}_\mu)$. Because $u_m$, $\mathcal{E}_\alpha$ and $\mathcal{H}_\mu$ are all external variables that cannot be changed by switching the polarization, then, $[\bar{C}_2\|\mathcal{T}] F_P = F_{-P}$ should lead to $-\alpha_{m\mu}u_m\mathcal{H}_\mu + (-\alpha_{\alpha\mu})\mathcal{E}_\alpha\mathcal{H}_\mu$ in the -P state. Therefore, the polarization switching through $[E\|\{I|t\}]$ with the simultaneous occurrence of the effects of $\mathcal{T}$ in real space through $[\bar{C}2\|\mathcal{T}]$ must switch the linear ME coefficients, including both the lattice ($\alpha_{m\mu}$) and electronic ($\alpha_{\alpha\mu}$) contributions. Then, if a static electric field is applied to the system, nonvolatile switchable weak ferromagnetism (i.e., $\mathcal{M} = \frac{\partial F}{\partial \mathcal{H}} \sim \alpha_{m\mu}u_m + \alpha_{\alpha\mu}\mathcal{E}_\alpha$) will be induced between the bipolar states shown schematically in **Fig. 3a** with the electric polarization switching path shown in the left of **Fig. 3a**. Last, since $[\bar{E}\|\mathcal{T}\{I|t\}] = [\bar{E}\|\mathcal{T}] \otimes [E\|\{I|t\}]$, the effects of $\mathcal{T}$ in real space disappear through the $[\bar{E}\|\mathcal{T}\{I|t\}]$-type polarization switching, where the ME coefficients cannot be switchable.

*Switchable ME coefficients and Generalization*—Next, we calculate the ferroelectrically switchable ME coefficients in Ca$_2$FeS$_3$Br. The ME response coefficients $\alpha_{yz}/\alpha_{zy}$ allowed by symmetry and fundamental to the DMI are studied using a first-principles scheme based on the electric-magnetic enthalpy $F = \alpha_{m\mu}u_m\mathcal{H}_\mu + \alpha_{m\alpha}u_m\mathcal{E}_\alpha$. This method simulates the linear ME response contributions from $\alpha_{m\mu}u_m$ and can give a reasonable estimation on the order of linear ME response [32,33], where an electric field along the out-of-plane direction was applied, thus inducing wFM along the *y* direction. Interestingly, we find a change of the sign of the $\alpha_{yz}$ for the -P2 and P states (**Fig. 3b**). The magnitude of the ME response coefficient $|\alpha_{yz}|$=0.65×10$^{-4}$ g.u.

(Gaussian units) is also comparable to the values of other materials, such as $Cr_2O_3$, $BiFeO_3$, and $LaFeO_3$/$YFeO_3$ short-period superlattices [32–34]. The energy barrier for switching the polarization is ~88 meV/f.u. for $Ca_2FeS_3Br$ with a computed polarization of 2.8 $\mu C/cm^2$, which indicates switching of ME coefficients may be realizable (**Fig. 3c**). Such property can be utilized for electric field controllable wFM in a nonvolatile manner. For example, a DC electric field along the out-of-plane direction induces wFM along the $y$ direction, then switching of the in-plane polarization through the OOR will switch the induced wFM owing to the change in sign of $\alpha_{yz}$ (**Fig. 3a**). Such kind of ME coupling was only realized in experiment in the PMN-PT/Terfenol-*D* heterostructure [35] and in the hexagonal $ErMnO_3$ with a magnetoelectric force microscopy technique that combines magnetic force microscopy with in situ modulating high electric fields [36], whose original theory was proposed in Lu$B$O$_3$ ($B$=Fe,Mn) [37]. The change in the sign of the linear ME coefficients in hexagonal manganites and ferrites relies on different kinds of 120° noncollinear spin orderings and the phases are not altermagnetic. Our proposed switchable ME coefficients may occur in a single multiferroic phase with the principal spin vectors forming simple collinear magnetic orderings at high temperature, e.g., the Néel temperature is 304 K in our Monte Carlo simulations for $Ca_2FeS_3Br$ (**Fig. 3d**).

By revisiting the band structures with SOCs between bipolar states in the other 16 altermagnetic multiferroics reported previously (**Table S7**), we further show that the symmetric bands dispersion and the switchable ME coefficients can always occur when the +P and -P states are connected by the spin space symmetry $[E\|\{I|t\}]$. This effect is independent of crystal family, extending beyond perovskite-like compounds. More interestingly, we find that there can be only two kinds of electric polarization switching paths, categorized by the spin space group symmetry $[E\|\{I|t\}]$ and $[\bar{E}\|\mathcal{T}\{I|t\}]$ (**Fig. 3a**). Both of them, however, are not guaranteed to simultaneously coexists in a compound. For example, there is only $[\bar{E}\|\mathcal{T}\{I|t\}]$-type polarization switching in $SrMn_2V_2O_8$ and there is only $[E\|\{I|t\}]$-type polarization switching in $Y_2Cu_2O_5$ (with the spinel-like structure). Such categorization is important: (1) they have mutually exclusive intrinsic ME couplings as demonstrated in our study, where $[\bar{E}\|\mathcal{T}\{I|t\}]$-type polarization switching cannot lead to the switchable ME coefficients, and $[E\|\{I|t\}]$-type polarization switching cannot have ferroelectric switchable altermagnetism. Therefore, among our 19 studied altermagnetic multiferroics, MnSe and $SrMn_2V_2O_8$ only exhibit spin-momentum locking phenomena whereas there is only switchable

ME coefficients in $Y_2Cu_2O_5$; the remaining 16 compounds can have both types of the polarization switching. (2) Our theoretical framework, together with the previously established theory of ferroelectrically switchable altermagnetism, demonstrates that altermagnetic materials inherently exhibit strong nonvolatile ME coupling. Further demonstrations of the switchable linear ME coefficients in room-temperature altermagnetic multiferroics, include, $BiFeO_3$ with stable 180° domain wall switching under strain and $(Ca,Sr)_{1.15}Tb_{1.85}Fe_2O_7$ whose linear ME coefficients have been measured at room temperature and 100 K, respectively [38–40].

Our theory extends beyond ferroelectrically switchable altermagnetism in that it is not restricted to altermagnetism. In many altermagnetic multiferroics, both kinds of switching paths can exist by symmetry (**Table S7**), where only the path having the lowest switching energy barrier can be realized. In our theory, we can control unambiguously the polarization switching path, thus, inducing exclusively the switchable ME coefficients. Here, a common chemical substitution method is introduced in altermagnetic multiferroics, that is, cation ordering, which we adopt to create two antiferromagnetically coupled sublattices with ferrimagnetism and suppress the $[\bar{E} \| \mathcal{T}\{I|t\}]$-type polarization switching. For example, if atoms 2 and 4 in **Fig. 3a** are changed to elements that are different from atoms 1 and 3, then polarization switching path through $[\bar{E} \| \mathcal{T}\{I|t\}]$ will be forbidden. We confirm this in cation ordered $LiNbO_3$-type multiferroics $BiFeCrO_6$ [32], $Zn_2FeOsO_6$ [41] and $Bi_2M ReO_6$ ($M$=Mn,Ni) [42] with rock-salt $B$ site orderings and ferrimagnetism (**Table S7**). Intriguingly, the switchable ME coefficients occur in their $R3$ polar ground state. This is because the $R3$ polar structure originates from $R3c$ by adopting rock-salt $B$ site orderings. The polarization switching route through $[\bar{E} \| \mathcal{T}\{I|t\}]$ in $R3c$ (e.g., $BiFeO_3$) is removed by the rock-salt $B$ site orderings. It should be noted that $BiFeCrO_6$ has a large transverse linear ME coefficient $\sim 14 \times 10^{-4}$ g.u. at 0 K and its thin film has Fe/Cr ordering and room-temperature multiferroicity as demonstrated in experiment [43]. In addition, $Zn_2FeOsO_6$ and $Bi_2M ReO_6$ ($M$=Mn,Ni) were predicted to be room-temperature multiferroics. Although the $[\bar{E} \| \mathcal{T}\{I|t\}]$-type polarization switching usually exists in altermagnetic multiferroics, there can be more ways, such as charge ordering, alloying, and vacancies, to suppress $[\bar{E} \| \mathcal{T}\{I|t\}]$-type polarization switching.

*Conclusions*—We formulated a structure-based design rule to identify the chemical compositions leading to polar ground states with switchable in-plane polarization in a series of heteroanionic $A_2BX_3X'$ ($X$=O, S and $X'$=N, Cl, Br) compounds. Through comprehensive electronic structure calculations, we established that members of the family include room-temperature multiferroic materials with hybrid improper ferroelectricity. A mechanism for changing the form of the linear ME coefficients by ferroelectricity was discovered, which is universal in the Type-I multiferroics. Last, our study discovered that altermagnetic multiferroics must have intrinsic ME coupling.

*Acknowledgments*—H.-M.Z. was supported by the National Natural Science Foundation of China (Grant Nos. 12347185), the Postdoctoral Fellowship Program of CPSF under Grant Number GZC20230443, and the Jiangsu Funding Program for Excellent Postdoctoral Talent. H.-M.Z., C.A.J. and X.-Z.L. were supported by the National Natural Science Foundation of China (NSFC) under Grant No. 12474081, the open research fund of Key Laboratory of Quantum Materials and Devices (Southeast University), Ministry of Education, the Start-up Research Fund of Southeast University. T.Z. and H.K. were supported by JSPS SUPRA-Ceramic Project (JP22H05143) and JST-ASPIRE (JPMJAP2408). J.M.R. was supported by the National Science Foundation (NSF) under DMR- 2413680. DFT calculations were performed through computational resources and staff contributions provided for the Quest high performance computing facility at Northwestern University which is jointly supported by the Office of the Provost, the Office for Research, and Northwestern University Information Technology. Part of the calculations were performed on high-performance computers, supported by the Big Data Computing Center of Southeast University.

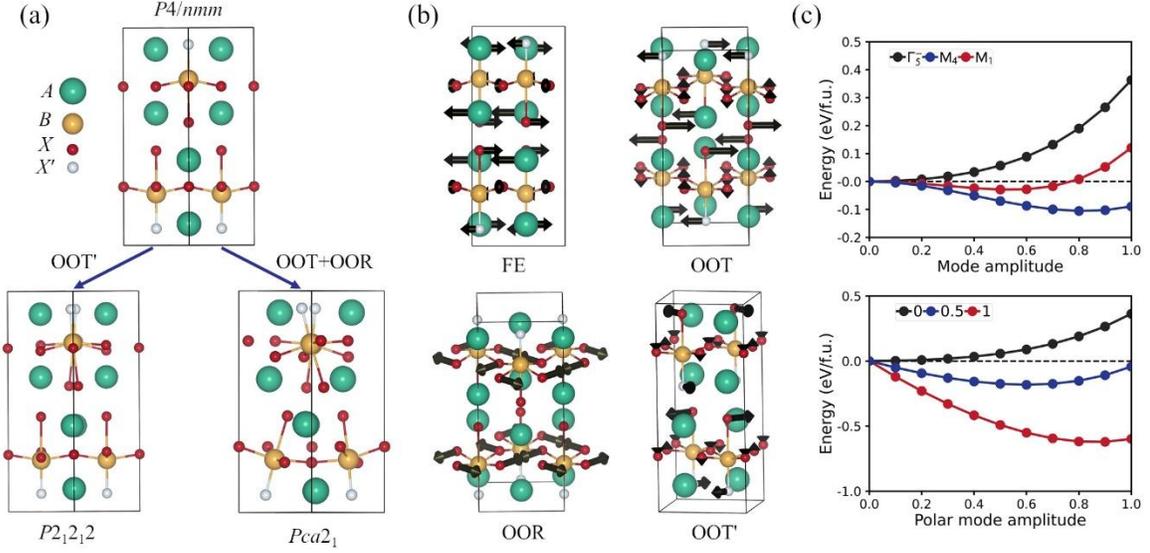

**FIG. 1**. (a) Structures of the ground state phases in the anion ordered $A_2BX_3X'$, where the antiparallel *trans* $X$-$B$-$X'$ arrangement of heterolepetic $BX_5X'$ octahedra between layers gives *P4/nmm* symmetry of the high-temperature phase. OOR, OOT and OOT' represents an oxygen octahedral rotation around an out-of-plane direction, and two kinds of oxygen octahedral tilts around an in-plane direction, respectively. (b) The main phonon modes describing the atomic displacements. FE: ferroelectric mode. (c) Upper panel: The energy changes with respect to the mode amplitude of $\Gamma_5^-$, $M_4$ and $M_1$ in $Ca_2FeS_3Br$. The mode amplitude is given relative to the amplitude of the mode in the ground state structure. Descriptions of the $\Gamma_5^-$ (FE mode), $M_4$ (OOR) and $M_1$ (OOT) displacements can be found in **Fig. 1b**. Bottom panel shows the appearance of the polarization in $Ca_2FeS_3Br$ with gradually increasing the magnitudes of both OOR and OOT, which is the feature of the hybrid improper ferroelectricity. "0", "0.5" and "1" indicate the magnitudes of the multiplications between the order parameters of OOR and OOT, i.e., $Q_{OOR} \times Q_{OOT}$.

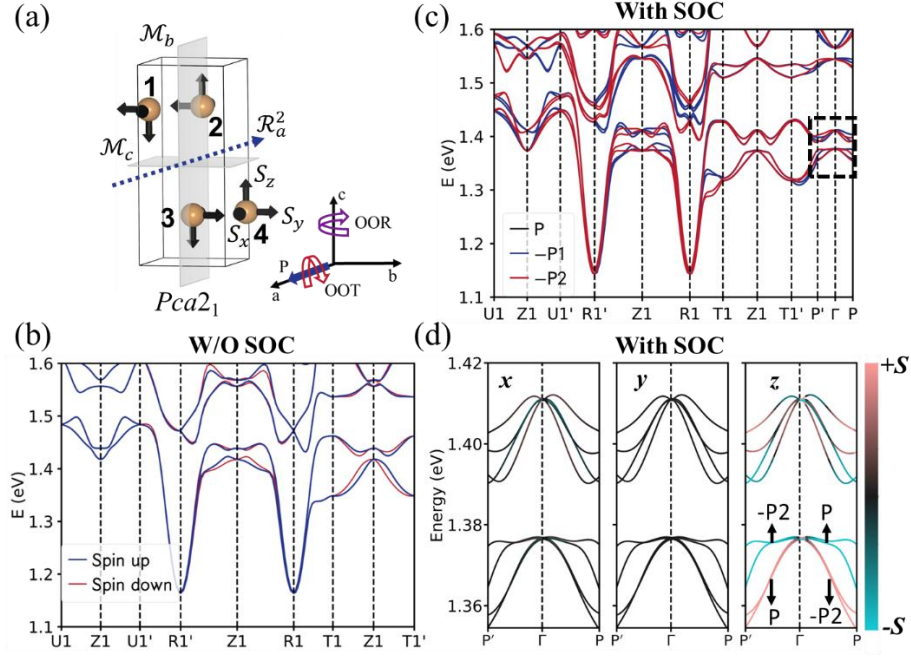

**FIG. 2**. (a) Crystal symmetry-adapted magnetic ground state structure. The magnetic space symmetries are shown. The black arrows indicate the spin directions $S=(S_x, S_y, S_z)$. The coordinate system with the directions of the main phonon modes in the structure are shown. The numbers indicate the different magnetic ions. (b) Electronic band structure of $Ca_2FeS_3Br$ without SOC at $k_z=\frac{2\pi}{5}$ plane, where the prime to the letter indicates the $k$ point has negative $k_x$ and $k_y$ components at $k_z=\frac{2\pi}{5}$ plane. (c) Electronic band structure around the CBM in the (P, OOT, OOR) (i.e., P), (-P, -OOT, OOP) (-P1) and (-P, OOT, -OOR) (-P2) states. (d) shows the electronic band structure enclosed with the dashed lines in (c) with the spin information for P and -P2 states. The spin along $z$ direction dominates in the bands dispersion along P'-Γ-P path.

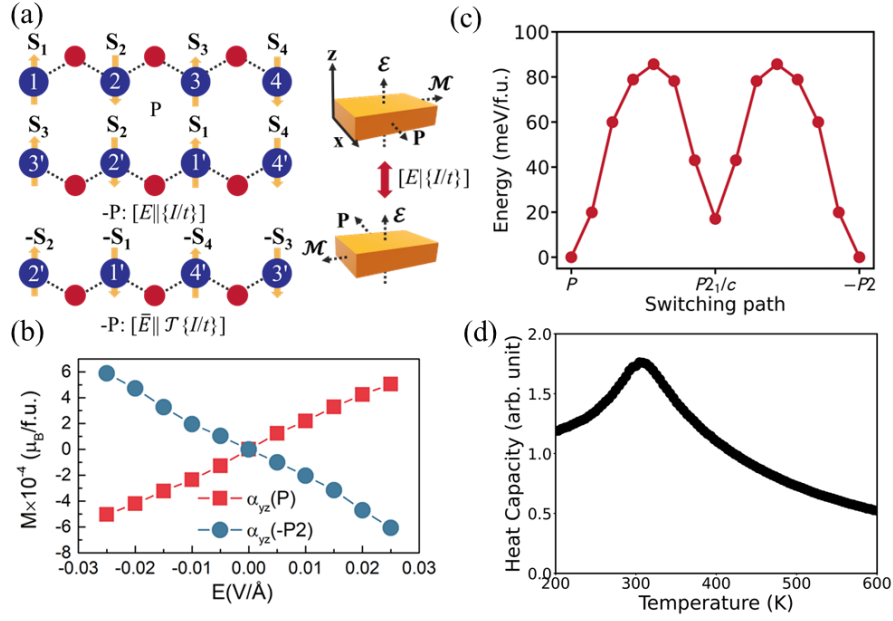

**FIG. 3**. (a) Schematic illustration of all the possible kinds of the 180° polarization switching paths in Type-I multiferroics (left) and the switching of ME response in $Pca2_1$ $Ca_2FeS_3Br$ (right). $\mathcal{M}$ is the wFM induced by the static electric field $\mathcal{E}$ through ME response. The position of the atom $i'$ ($x'$, $y'$, $z'$) in the -P structure is ($-x+t$, $-y+t$, $-z+t$) obtained by applying $\{I|t\}$ to the atom $i$ in the P structure ($i$=1-4). The associated change of the spin ($S_i$) direction on the atom $i$ is derived from DMI energies by applying $[E\|\{I|t\}]$ and $[\bar{E}\|\mathcal{T}\{I|t\}]$. (b) The change in wFM along $y$ direction in the P and -P2 states by varying the electric field along the $z$ direction, i.e., the linear ME response coefficients $\alpha_{yz}$ in the two states. (c) Polarization switching barrier for $Pca2_1$ $Ca_2FeS_3Br$ through a centrosymmetric intermediate ($P2_1/c$). (d) Calculated Néel temperatures for $Ca_2FeS_3Br$ obtained from Monte Carlo simulations. Up to next-nearest-neighboring symmetric spin exchange interactions, nearest-neighboring antisymmetric Dzyaloshinsky-Moriya interactions, and single-ion anisotropy are included in the simulations [27].